\documentstyle[12pt]{article}
\begin{document}
\title{\vspace*{-1.8cm}
Combining Information from $B \to \pi \pi $ and $B \to \pi \rho, \pi \omega $ 
Decays }
\author{
{M. Sowa and P. \.Zenczykowski $^*$ }\\
\\
{\em Dept. of Theoretical Physics},
{\em Institute of Nuclear Physics}\\
{\em Polish Academy of Sciences}\\
{\em Radzikowskiego 152,
31-342 Krak\'ow, Poland}\\
}
\maketitle
\begin{abstract}

We consider the $B \to \pi \pi $ and $B \to \pi \rho, \pi \omega $ decays
alongside each other,
taking into account the contributions from all individual penguin amplitudes   
generated by the internal $t$, $c$, and $u$ quarks.
We {{argue}} that three ratios of penguin amplitudes, each for a different
internal quark, formed by dividing the
individual penguin amplitude in $B \to \pi \pi$ by the corresponding
amplitude in $B \to \pi \rho, \pi \omega $, 
should be equal. 
We study the implications of the {{assumed}} existence of this connection
between $B\to \pi \pi$ and $B \to \pi \rho, \pi \omega $.
First, {{accepting}} that in the $B \to \pi \pi$ decays the ratio $C/T$ 
of the colour-suppressed 
factorization amplitude $C$ to the tree
factorization amplitude $T$ is negligible,  we determine the
ratio of individual penguin amplitudes.
 Then, from the $B \to \pi \rho, \pi \omega$ data, we extract the  
 effective (i.e. possibly containing some penguin terms) tree and 
 the effective colour-suppressed 
 amplitudes relevant for these processes, and the
corresponding solutions for the factorization
amplitudes.
Finally, we argue that the $C/T$ ratio in $B \to \pi \pi $ should be
identical to its counterpart in
 $B \to \pi \rho, \pi \omega$ (relevant for pion emission from
the decaying $b$ quark).
This constraint permits the determination of $C/T$ and of 
other amplitude ratios 
directly from the data. Although the $|C/T|$ ratio extracted
from the available data still carries a substantial error, it is
consistent with the expected value of 0.25 - 0.5.

\end{abstract}
$^*$ E-mail:
zenczyko@iblis.ifj.edu.pl
\newpage

\section{Introduction}

Decays of $B$ mesons to charmless final states provide us with a lot of 
information
concerning the phases of the Cabibbo-Kobayashi-Maskawa matrix. 
Consequently, these decays
have been the subject of numerous studies.
The problem is to extract relevant information from the data, since any such
procedure involves serious
uncertainties resulting from 
our poor knowledge of the effects of strong interactions.
Many papers have been devoted to the analysis of $B \to PP$
decays (with $P$ denoting a pseudoscalar meson), in the hope that the abundance
of data will permit the determination of several weak and strong parameters 
involved.

In recent papers \cite{BurasFleischer} it was shown that the data on 
$B \to \pi \pi$ decays require the
presence of important nonfactorizable corrections
and hadronic interference effects if
the SM value of $\gamma \approx 65^o$
and the  CP-averaged $B_d \to \pi ^0 \pi^0$ 
 branching ratios recently measured by BaBar and Belle
\cite{bb} are used.
The authors of ref. \cite{BurasFleischer} show in a theoretically clean way
that  the effective tree
and colour-suppressed amplitudes $\tilde{T}$ and $\tilde{C}$ 
governing the $ B \to \pi \pi $ decays
are roughly equal in absolute magnitudes,
thus contradicting the expectation that the latter amplitude
should be substantially suppressed.
In order to conform to this expectation, one has to admit that the
corrections 
in the effective
amplitudes $\tilde{T}$ and $\tilde{C}$ 
arising from the usually neglected penguin contributions
are substantial (see also \cite{Alietal} and \cite{BURAS2004end}).
Thus, hadronic-level effects appear to invalidate the naive expectation of the 
factorization prescription.
Indeed, final-state strong
interaction effects should contribute to the redefinition of the original
quark-diagram amplitudes, thus generating the usually neglected penguin
contributions referred to above (see refs. \cite{LZ20022003,Z2004}).

In general, better extraction of the relevant strong and weak parameters 
requires considering a larger body of data. Recently, several analyses
appeared in which extraction of the angle $\gamma $ of the unitarity triangle
(UT) was attempted from a fit to all currently available data on
$B \to PP$  
or $B \to PV$ decays (see e.g. \cite{Z2004,Cot,GR2004}). 
Such analyses in the $B \to PP$ and $B \to PV$ sectors are usually performed
separately from one another. On the other hand, arguments may be given that
some of the parameters, introduced in these two sectors to take account of the
effects of strong interactions, are actually related. By combining
in a single analysis the information
from both sectors, one could then hopefully get an additional handle on  
the previously undetermined parameters.

In this paper, we analyse the decays of $B$ to $\pi \pi$, $ \pi \rho $, and 
$\pi \omega$
in an approach modeled on ref. \cite{BurasFleischer}. 
The data on these decays
 should provide sufficient 
information on which to base the
analysis and comparison of the size of all factorization 
amplitudes involved.
In principle, there is no need here to use the data on strangeness-changing
two-body decays of $B$ mesons, a welcome feature since $B\to \pi K$
decays exhibit various puzzles (in addition to substantial 
contributions from electroweak penguins),
 as analysed in \cite{BurasFleischer}.
In practice, however, the data on $B \to \pi \pi$ are
still not good enough, and we find it necessary 
to determine the magnitude of  penguin amplitude $P$
from $B^+ \to \pi^+ K^0$.
This transition and the related $B^+ \to \pi^+ K^{*0}$ decay constitute 
the only places 
wherefrom information from the strangeness-changing
sector enters into our analysis. The knowledge of $|P|$ permits the extraction of several 
$B \to \pi \rho, \pi \omega$ parameters from the data
and the determination of amplitude
ratios, such as e.g. $\tilde{T}/P$ etc. (i.e. it gives the size of various
amplitudes in units of $|P|$).

In Section 2, 
{{we set out our notation following refs \cite{BurasFleischer,BURAS2004end}
and,  using the Summer 2004 data given by the HFAG 
\cite{HFAGsummer2004} we repeat this part of the
$B\to \pi \pi $ analysis of ref. \cite{BURAS2004end} which is
 relevant for our purposes.}}
 We then present a simple yet illuminating formula which
expresses the ratio of penguin amplitudes involving loops with 
different quarks in terms of the ratio
of factorization amplitudes $C/T$ and the parameters extracted from the data.  
In Section 3, using the data on $B^+ \to \pi^+ K^{(*)0}$ decays, 
we determine the absolute magnitude of penguin amplitudes 
{{and formulate our main assumption concerning their ratios}}.
In Section 4, we proceed with an analysis of the $B \to \pi \rho$
and $B \to \pi \omega $ decays.
We find that there are two acceptable sets of solutions for the effective
colour-suppressed and tree amplitudes.
Assuming that the ratio of penguin amplitudes involving loops with different
quarks is well approximated by the formula of Section 2 with $C/T=0$, we
determine the tree and colour-suppressed factorization amplitudes in
$B \to \pi \rho, \pi \omega$.
In Section 5, we give
a $B \to PV$ analog of the formula given in Section 2, 
expressing the ratio of penguin amplitudes in terms of the ratio of
factorization amplitudes in the $B \to \pi \rho, \pi \omega$ sector.
We combine this formula with its counterpart from Section 2 and
directly from the data determine both
$C/T$ and the ratio of penguin amplitudes involving loops with 
different quarks.
Our conclusions are contained in
Section 5.

\section{Decays $B \to \pi \pi $}
\label{section2}

The $B \to \pi \pi$ amplitudes 
may be expressed in terms of amplitudes
$P$ (penguin), $\tilde{T}$ (effective tree), and $\tilde{C}$
(effective colour-suppressed): 
\begin{eqnarray}
\label{Btopipione}
\sqrt{2} A(B^+ \to \pi ^+ \pi ^0)&=&-[\tilde{T}+\tilde{C}]\\
\label{Btopipitwo}
A(B^0_d \to \pi ^+ \pi ^-)&=&-[\tilde{T}+P] \\
\label{Btopipithree}
\sqrt{2} A(B^0_d \to \pi ^0 \pi ^0)&=&-[\tilde{C}-P], 
\end{eqnarray}
with:
\begin{eqnarray}
\label{Btopipi1}
P \equiv P_c \phantom{xxxxx}&=& A \lambda ^3{\cal{P}}_{tc}
\\
\label{Btopipi2}
\tilde{T}\equiv e^{i \gamma} (T-R_b P_u)&=&
A \lambda ^3 R_b e^{i \gamma}({\cal{T}}-{\cal{P}}_{tu}) \\
\label{Btopipi3}
\tilde{C}\equiv e^{i \gamma} (C+R_b P_u)&=&
A \lambda ^3 R_b e^{i \gamma}({\cal{C}}+{\cal{P}}_{tu}),
\end{eqnarray}
where the rightmost forms involve the definitions of refs 
\cite{BurasFleischer,BURAS2004end},
from which we omitted the contributions due to the exchange amplitudes.
{{In principle, the latter 
could be included by a mere redefinition of $P_{tu}$.
However, the general idea of this paper would then require an analogous
redefiniton of penguin amplitudes in $B \to \pi \rho, \pi \omega$, with the
relative size of penguin and exchange amplitudes unlikely to be the same as in
$B \to \pi \pi$. Only if exchange amplitudes are neglected (as
usually done and assumed hereafter) and $P_{tu}$ represents just the
 difference between the top- and up-
penguins, our approach does not need additional parameters.}}

In Eqs (\ref{Btopipi1}-\ref{Btopipi2}), $A=0.83 \pm 0.02$, $\lambda = 0.224 \pm 0.0036$, $R_b=0.37\pm0.04$, 
and  $\gamma $
parametrize the Cabibbo-Kobayashi-Maskawa (CKM) matrix. 
$T \equiv A\lambda^3R_b \cal{T}$ and 
$C \equiv A\lambda^3R_b \cal{C}$ involve
the strong amplitudes of colour-allowed and colour-suppressed
tree diagrams. Finally,
$P_q\equiv -\lambda ^{(d)}_c {\cal{P}}_{tq}= A\lambda^3{\cal{P}}_{tq}\equiv 
A\lambda^3({\cal{P}}_t-{\cal{P}}_q)$ with ${\cal{P}}_k$
describing penguin strong amplitudes corresponding to 
internal k-quark exchanges
($k \in \{t,c,u\} $), i.e. with the full penguin amplitude
given by $\lambda ^{(d)}_u {\cal{P}}_u + \lambda ^{(d)}_c {\cal{P}}_c +
\lambda ^{(d)}_t {\cal{P}}_t$, where $\lambda ^{(k)}_q = V_{qk}V^*_{qb}$, 
and $V$ is the CKM matrix.
For flavour-symmetric final-state interactions (FSI), 
the above formulae encompass
all elastic and inelastic FSI with the exception of those 
represented by crossed diagrams
(see \cite{LZ20022003,Z2004}).
When the latter are taken into account, the
amplitudes $T$ and $C$ become mixtures of the colour-allowed and 
colour-suppressed factorization amplitudes \cite{LZ20022003,Z2004}.

\subsection{Extraction of hadronic parameters}
Introducing the hadronic parameters {{of refs
\cite{BurasFleischer,BURAS2004end}}}:
\begin{eqnarray}
\label{dtheta}
d e^{i\theta} \equiv - e^{i
\gamma}P/\tilde{T}&=&\phantom{xx}-|P/\tilde{T}|e^{-i\delta_{\tilde{T}}} 
=-P_c/(T-R_bP_u)\\
\label{xDelta}
x e^{i \Delta} \equiv
\tilde{C}/\tilde{T}\phantom{xx}&=
&|\tilde{C}/\tilde{T}|e^{i(\delta_{\tilde{C}}-\delta_{\tilde{T}})} 
=(C+R_bP_u)/(T-R_bP_u)
\end{eqnarray}
where $\delta_{\tilde{T}},\delta_{\tilde{C}}$ denote strong phases
(in the convention in which
 the strong phase of the penguin amplitude $P_c$ is assumed zero),
one can derive the following formulae relating the parameters just introduced
to the branching ratios ${\cal{B}}$ and asymmetries in $B \to \pi \pi$ decays
\cite{BurasFleischer}:
\begin{eqnarray}
\label{BR1}
R^{\pi \pi}_{+-}&=&\frac{1+2x \cos \Delta + x^2}{1-2d \cos \theta \cos \gamma
+d^2}\\
\label{BR2}
R^{\pi \pi}_{00}&=&\frac{d^2+2dx 
\cos (\Delta -\theta)\cos \gamma + x^2}{1-2d \cos \theta \cos \gamma
+d^2}\\
\label{Adir}
A^{{\rm dir}}_{\pi ^+ \pi ^-}&=
&-\frac{2d \sin \theta \sin \gamma}{1-2d \cos \theta \cos \gamma
+d^2}\\
\label{Amix}
A^{{\rm mix}}_{\pi ^+ \pi ^-}&=
&\frac{\sin (2 \beta + 2 \gamma) - 
2d \cos \theta \sin (2\beta +\gamma)+ d^2 \sin (2\beta )}
{1-2d \cos \theta \cos \gamma+d^2},
\end{eqnarray}
with \cite{HFAGsummer2004}
\begin{eqnarray}
\label{dataRpippim}
R^{\pi \pi}_{+-}&\equiv & 2 ~\frac{{\cal{B}}(B^{\pm}\to \pi ^{\pm}\pi ^0)}
{{\cal{B}}(B_d \to \pi ^+ \pi ^-)}~\frac{\tau_{B^0_d}}{\tau_{B^+}}=2.20\pm 0.31\\
\label{dataRpi0pi0}
R^{\pi \pi}_{00}&\equiv & 2 ~\frac{{\cal{B}}(B_d\to \pi ^0 \pi ^0)}
{{\cal{B}}(B_d \to \pi ^+ \pi ^-)}=0.66 \pm 0.13\\
\label{datadir}
A^{{\rm dir}}_{\pi ^+ \pi ^-}&=&-0.37 \pm 0.11 \\
\label{datamix}
A^{{\rm mix}}_{\pi ^+ \pi ^-}&=&+0.61\pm 0.14,
\end{eqnarray}
where, {{as in original papers \cite{BurasFleischer,BURAS2004end}}}, 
the asymmetries are estimated as weighted averages, in spite of
the BaBar and Belle results still not being fully consistent.
{{Since these averages have not changed much in comparison to
information available before Summer 2004, we believe that performing the
analysis of this paper for the BaBar and Belle asymmetries separately is
unwarranted.}}

Assuming $\beta =24^o$ and $\gamma =65^o$, one can determine $d$ and $\theta$
using  Eqs (\ref{Adir},\ref{Amix}) and the experimental values of the
 asymmetries from Eqs (\ref{datadir},\ref{datamix})
\cite{BurasFleischer}.
Two solutions for $(d,\theta)$ are obtained, 
of which refs \cite{BurasFleischer,BURAS2004end} 
accept only the one with small $d$, 
the other one (with $d \approx 4.6 $) being excluded as it
leads to complex solutions.

With the updated values of asymmetry averages, {{we obtain}}
\begin{eqnarray}
\label{dexp}
d&=& 0.52^{+0.22}_{-0.14} \\
\label{thetaexp}
\theta&=& +(141^{+11}_{-12})^o.
\end{eqnarray}
In \cite{BurasFleischer} (\cite{BURAS2004end})
the corresponding values were: $d=0.49^{+0.33}_{-0.21}$ ($0.51^{+0.26}_{-0.20}$),
and $\theta=+(137^{+19}_{-23})^o$ ($+(140^{+14}_{-18})^o$).
We determine the errors as in \cite{BurasFleischer}, i.e.
the errors associated with the specific input parameter 
(here: asymmetry or an $R^{\pi \pi}$
ratio ) are estimated by varying its value within $1\sigma $ while keeping
the other input parameters at their central values, with the individual
errors thus obtained subsequently added in quadrature.

The solution of Eqs (\ref{BR1},\ref{BR2},\ref{dataRpippim},\ref{dataRpi0pi0}) 
yields
\begin{eqnarray}
\label{xexp}
x&=&\ 1.13 ^{+0.16}_{-0.15}\\
\label{Deltaexp}
\Delta &=& -(55^{+17}_{-26})^o,
\end{eqnarray}
to be compared with $x=1.22^{+0.25}_{-0.21}$ ($1.13^{+0.17}_{-0.16}$), 
and $\Delta = -(71^{+19}_{-25})^o$ ($-(57^{+20}_{-30})^o$) 
in refs \cite{BurasFleischer} {{and \cite{BURAS2004end} respectively}}.
As in \cite{BurasFleischer}, 
we have discarded the second solution for $x$ and $\Delta$ 
(with $x \approx 0.96$, and $\Delta \approx +33^o $), since  
the $A_{CP}(B^+\to K^0\pi^+)$ asymmetry it yields is of the order 
of $0.2$ (using Eq.(\ref{ACPpiplusK0}) below), too large when compared with the experimental data 
of \cite{HFAGsummer2004}: $A_{CP}^{exp}(B^+\to K^0\pi^+)=-0.02 \pm 0.034 $.
A similar argument was originally used in \cite{BurasFleischer} in connection with 
$A_{CP}(B^+\to K^+\pi^0) $.
The experimental value of the $A_{CP}(B^+\to K^0\pi^+)$ asymmetry puts
a much stronger bound on the error in $\Delta-\theta$ (see below).

\subsection{Ratio of penguin amplitudes}
From Eqs (\ref{dtheta},\ref{xDelta}) one derives:
\begin{equation}
\label{PctoPu}
\frac{P_c}{P_u}=\frac{{\cal{P}}_{tc}}{{\cal{P}}_{tu}}=
-(1+\frac{C}{T})\frac{R_b~de^{i\theta}}{xe^{i\Delta}-\frac{C}{T}}
\end{equation} 
Since the value of $|C/T|$ is expected to be of the order of $0.25$ only, 
a good estimate of the ratio of penguin terms should be obtained by setting
$C/T=0$ above:
\begin{equation}
\label{PctoPuapprox}
\frac{P_c}{P_u}\approx -\frac{R_b~d~e^{i\theta}}{xe^{i\Delta}} 
\approx (0.17^{+0.08}_{-0.05} ) ~e^{i (16^{+21}_{-28})^o}
\end{equation}
with
\begin{equation}
\label{Deltamintheta1}
\Delta-\theta = (+164^{+28}_{-21})^o
\end{equation}
where from now on we treat the errors of $d$, $\theta$, $x$, and $\Delta$
as independent, adding in quadrature the corresponding errors to
the moduli and phases {{of quantities depending on these parameters.}}
{{ Since the errors on $d$, $\theta$, $x$ and $\Delta$ are actually
inter-related, another possible treatment of errors would be to vary the
original four parameters ($R^{\pi \pi}_{+-}$, $R^{\pi \pi}_{00}$,
$A^{{\rm dir}}_{\pi ^+ \pi ^-}$, $A^{{\rm mix}}_{\pi ^+ \pi ^-}$) within their
error bars. We have not chosen this more involved
route since possible errors stemming from the lack of consistency among
the BaBar and Belle asymmetry results for $A^{{\rm dir}}_{\pi ^+ \pi ^-}$, 
$A^{{\rm mix}}_{\pi ^+ \pi ^-}$  might render its
higher quality questionable. Error analysis of
this paper,  with $R^{\pi \pi}_{+-}$, $R^{\pi \pi}_{00}$,
$A^{{\rm dir}}_{\pi ^+ \pi ^-}$, $A^{{\rm mix}}_{\pi ^+ \pi ^-}$ varied within
their error bars, could be repeated
when the input data are under better control. }}

From Eq. (\ref{dtheta}) one further obtains :
\begin{equation}
\label{TtoPcvalue}
\frac{T}{P_c}=R_b\frac{P_u}{P_c}-\frac{1}{d}e^{-i\theta}\approx 
-\frac{1+x~e^{i\Delta}}{d~e^{i\theta}}\approx (3.6^{+1.4}_{-1.1}) 
~e^{i~(10 ^{+18}_{-15})^o}
\end{equation}
and 
\begin{equation}
\label{TtoPuRbvalue}
\frac{T}{R_bP_u}\approx 1+ \frac{1}{x}e^{-i~\Delta} 
\approx (1.68 ^{+0.19}_{-0.18})~ e^{i~ (26 ^{+8}_{-12}) ^o}
\end{equation}
which demonstrates that the size of the penguin correction term 
($R_bP_u$) with respect to the
tree amplitude $T$ is large, as discussed in \cite{BurasFleischer}.

When small nonzero values of $C/T$ are admitted, the ratio $P_c/P_u$ receives
a correction term
\begin{equation}
\delta (P_c/P_u) \approx \frac{R_b~d~e^{i \theta}}{x ~e^{i\Delta}}
\left ( \frac{1}{x~e^{i\Delta}}-1 \right )\frac{C}{T}\approx 0.15
e^{-i~40^o}\frac{C}{T}
\end{equation}
whose inclusion, as shown by the r.h.s. above 
(obtained by inserting the central values of  $x$, $d$, etc.), 
increases the errors in our estimate of $|P_c/P_u|$ 
given in Eq. (\ref{PctoPuapprox}) by some 20\% (for $|C/T| \approx 0.2$).

\section{Size of penguin amplitudes} 

Given the current errors in the determination of the $|\tilde{T}/{P}|$ ratio
(c.f. Eq. (\ref{dexp})),
further information on the size of amplitudes considered in
Section {\ref{section2}} is best obtained from the decays of 
$B^+\to \pi ^+ K^{0}$ ($B^-\to \pi ^- \bar{K}^{0}$) and 
$B^+ \to \pi ^+ K^{*0} $ ($B^- \to \pi ^- \bar{K}^{*0} $)
as these are expressed
in terms of penguin amplitudes alone:
\begin{eqnarray}
A(B^+ \to \pi ^+ K^0)& = & \tilde{P}'\\
A(B^+ \to \pi ^+ K^{*0})& =& \tilde{P}'_P,
\end{eqnarray}
with
\begin{eqnarray}
\label{Pprim}
\tilde{P}'&=
&-\lambda ^{(s)}_u{{\cal{P}}_{tu}}-\lambda^{(s)}_c{{\cal{P}}_{tc}}\\
\label{PprimP}
\tilde{P}'_P&=
&-\lambda ^{(s)}_u{{\cal{P}}_{P,tu}}-\lambda^{(s)}_c{{\cal{P}}_{P,tc}},
\end{eqnarray}
where the subscript $_P$ ($_V$) for $B \to PV $ amplitudes denotes amplitudes
in which the spectator quark ends in the final  $P$ ($V$) meson.
{{As already remarked in the Introduction this constitutes an implicit use of
SU(3), at present necessary, given the size of errors in the $\pi \pi$ sector.
In fact, knowing the absolute size of penguin amplitudes well
 is important in the formulas of
Section 4 (see e.g. Eqs (\ref{tanThetaVkV}-\ref{lV})), 
with the route via $K\pi$ etc. decays
yielding much smaller errors. 
With the size of penguin amplitude depending upon the SU(3) assumption in question, the latter affects 
also the extracted values of tree and colour-suppressed amplitudes.}}

\subsection{ \mbox {\boldmath ${B^+ \to \pi ^+ K^{0}}$}}
Let us consider the $B^+ \to \pi ^+ K^ {0}$ and $B^- \to \pi ^- \bar{K}^ {0}$
decays.
We introduce $P'_c$ (the analog of $P_c$):
\begin{equation}
\label{Pcprim}
P'_c=-\lambda ^{(s)}_c {\cal{P}}_{tc}=
-A \lambda ^2(1-\lambda ^2/2) {\cal{P}}_{tc}.
\end{equation}
Using Eq. (\ref{PctoPuapprox}) for the ratio of ${\cal{P}}_{tu}/{\cal{P}}_{tc}$
one obtains from Eqs (\ref{Pprim},\ref{Pcprim}):
\begin{equation}
\label{Pcprimcorrected}
\tilde{P}'=P'_c \left[ 1+\epsilon R_b \frac{{\cal{P}}_{tu}}{{\cal{P}}_{tc}} \right] 
\approx P'_c\left[ 
1-\epsilon\frac{x}{d}\frac{e^{i\Delta}}{e^{i\theta}}
e^{i\gamma}
\right],
\end{equation}
where $\epsilon = \lambda^2/(1-\lambda ^2) = 0.05 $.

The CP-asymmetry $A_{CP}(B^+\to \pi^+ K^0)$ is approximately
\begin{equation}
\label{ACPpiplusK0}
A_{CP}(B^+ \to \pi ^+ K^0) \approx 
\frac{-2 \epsilon \frac{x}{d}\sin (\Delta - \theta)\sin
\gamma}{1-2 \epsilon \frac{x}{d}\cos (\Delta - \theta)\cos
\gamma}
\end{equation}
which, together with its experimental value of $-0.02\pm 0.034$, 
and $x/d=2.17 ^{+0.86}_{-0.70}$,
points towards 
$\Delta -\theta \approx (+5 ^{+10}_{-9})^o$ or $(+174 \pm 11)^o$,
significantly improving upon the value of Eq. (\ref{Deltamintheta1}).
Thus, the experimental value of the $A_{CP}(B^+\to \pi^+ K^0)$ asymmetry
forces $\Delta - \theta$ to be close to $0^o$ or $180^o$, rejecting the 
solution with $\Delta \approx 33^o$ (for which $\Delta -\theta \approx -108 ^o$).
Consequently,  from now on, whenever  
only $\Delta -\theta$ appears instead of $\Delta$ and $\theta$, 
we shall use the average of the two determinations:
\begin{equation}
\label{Deltamintheta2}
\Delta - \theta =+(173 \pm 10)^o.
\end{equation}
Thus, the right hand side of Eq. (\ref{PctoPuapprox}) is replaced with
\begin{equation}
\label{PctoPuapproxnew}
\frac{P_c}{P_u} 
\approx (0.17^{+0.08}_{-0.05} ) ~e^{i (7{\pm 10})^o}.
\end{equation}

The CP-averaged branching ratio of $B^+ \to \pi^+ K^0$ is approximately:
\begin{equation}
\label{CPaverBRBplustopiK}
{\cal{B}}(B^{\pm} \to \pi^{\pm} K^0(\bar{K^0}))\approx
|P'_c|^2
\left[
1-2\epsilon \frac{x}{d}
\cos (\Delta - \theta) \cos \gamma
\right].
\end{equation}
Using the experimental value of this branching ratio: 
${\cal{B}}_{exp}(B^{\pm} \to \pi^{\pm} K^0(\bar{K^0}))=24.1 \pm 1.3$
(in units of $10^{-6}$), 
$x/d=2.17 ^{+0.86}_{-0.70}$, and $\Delta - \theta = +(173 \pm 10)^o$, 
one finds:
\begin{equation}
|P'_c|^2 \approx 22.1 ^{+1.3} _{-1.4}
\end{equation}
i.e.
\begin{equation}
\label{Pprimc}
P'_c=-4.70 ^{+0.14}_{-0.15},
\end{equation}
where the strong phase of $P'_c$ (originating from ${\cal{P}}_{tc}$) 
is assumed zero by SU(3) symmetry with $P_c$.
One then finds
\begin{equation}
\label{Pcvalue}
P_c=-\frac{\lambda}{1-\lambda^2/2}P'_c=-(0.230 \pm 0.004) P'_c = 1.08 \pm 0.04
\end{equation}

\subsection{\mbox{\boldmath $B^+ \to \pi ^+ K^{*0}$}}
For the description of 
$B^+ \to \pi ^+ K^ {*0}$ and $B^- \to \pi ^- \bar{K}^ {*0}$
decays, in analogy to $P'_c$ and $P_c$, we introduce:
\begin{equation}
\label{PPprimc}
P'_{P,c}=-\lambda ^{(s)}_c {\cal{P}}_{P,tc}=
-A \lambda ^2(1-\lambda ^2/2) {\cal{P}}_{P,tc}.
\end{equation}


{{To proceed we assume that}}
\begin{equation}
\label{universalpenguins}
\frac{{\cal{P}}_{P,tu}}{{\cal{P}}_{P,tc}}=\frac{{\cal{P}}_{tu}}{{\cal{P}}_{tc}}.
\end{equation}

{{The above equality follows if one accepts that the formation of
the final $PP$ or $PV$ pair is independent of penguin 
transition occurring before that formation takes place. 
This should be so if the intermediate
$s\bar{u}$ state does not remember how it was produced. The relevant
penguin-induced decay amplitude becomes then a product of a penguin term and the
amplitude describing the formation of the final state,
 with the latter amplitude cancelling out in the ratios 
${{\cal{P}}_{P,tu}}/{{\cal{P}}_{P,tc}}$ 
and ${{\cal{P}}_{tu}}/{{\cal{P}}_{tc}}$.
The above assumption is a crucial assumption upon which the rest of this paper
is based. }}

{{Assuming Eq. (\ref{universalpenguins}),}}
an analog of Eq.(\ref{Pcprimcorrected}) follows:
\begin{equation}
\tilde{P}'_P=P'_{P,c} \left[ 1+\epsilon R_b \frac{{\cal{P}}_{tu}}{{\cal{P}}_{tc}} \right] 
\approx P'_{P,c}\left[ 
1-\epsilon\frac{x}{d}\frac{e^{i\Delta}}{e^{i\theta}}
e^{i\gamma}
\right].
\end{equation} 
Consequently, from the experimental branching ratio of
${\cal{B}}_{exp}(B^{\pm} \to \pi^{\pm} K^{*0}(\bar{K^{*0}}))=9.76 ^{+1.16}_{-1.22}$
and using the analog of Eq. (\ref{CPaverBRBplustopiK}):
\begin{equation}
\label{CPaverBRBplustopiKstar}
{\cal{B}}(B^{\pm} \to \pi^{\pm} K^{*0}(\bar{K^{*0}}))\approx
|P'_{P,c}|^2
\left[
1-2\epsilon \frac{x}{d}
\cos (\Delta - \theta) \cos \gamma
\right]
\end{equation}
one determines that
\begin{equation}
\label{PPprimcvalue}
\left| \frac{P'_{P,c}}{P'_c}\right| ^2=
\frac{{\cal{B}}_{exp}(B^{\pm} \to \pi^{\pm} 
K^{*0}(\bar{K^{*0}}))}{
{\cal{B}}_{exp}(B^{\pm} \to \pi^{\pm} K^0(\bar{K^0}))}
\end{equation}
Introducing the ratio $\xi \equiv P_{P,c}/P_c$ 
so that the $B \to PV $ amplitudes may be later expressed in units of  
the $B \to PP$ penguin amplitude $P_c$ 
(to be compared with Eqs (\ref{TtoPcvalue},\ref{TtoPuRbvalue})),
one then finds that
\begin{equation}
\label{ksi}
\xi = P_{P,c}/P_c ={\cal{P}}_{P,tc}/{\cal{P}}_{tc}
={\cal{P}}_{P,tu}/{\cal{P}}_{tu}=P'_{P,c}/P'_c = 0.64 \pm 0.04
\end{equation}
with
\begin{equation}
P'_{P,c}=\xi P'_c=-2.99 \pm 0.21,
\end{equation}
where we adopted the convention of a vanishing strong phase for the $P'_{P,c}$
penguin amplitude (${\cal{P}}_{P,tc}$).
One then finds
\begin{equation}
\label{PPcvalue}
P_{P,c}=-(0.230 \pm 0.004) P'_{P,c} = 0.69 ^{+0.05}_{-0.05}.
\end{equation}

\section{Decays \mbox{\boldmath $B \to \pi \rho, \pi \omega $}}
Besides the amplitudes $P_{P(V)}$ ($P_{P(V),c}$) already considered 
in the previous section, 
strangeness-conserving decays of $B$ mesons 
into a pseudoscalar-vector meson pair
introduce several further amplitudes: $T_{P(V)}$ (tree), $C_{P(V)}$
(colour-suppressed), $S_{P(V)}$ (singlet penguin), etc., of which
$T_{P(V)},P_{P(V)},C_{P(V)}$ are considered to be the dominant ones.
The Zweig rule suggests that $S_P$ should be negligible.
On the other hand, $S_V$ does not need to be, 
in analogy to the situation in the
$B \to PP $ sector where decays $B \to K \eta (\eta ')$ seem to
indicate the non-negligible size of $S'$ \cite{Lipkin,RosnerLipkin}.
Since we want 
to restrict our analysis to a group of decays akin to $B \to \pi \pi$,
i.e. not
involving singlet penguin amplitudes, we are left with the following 
$B^0_d,B^+$ decay
channels to be considered:
$\pi ^+ \rho ^-$, $\pi ^- \rho ^+$, $\pi ^0 \rho ^+$, $\pi ^+ \rho ^0$,
$\pi ^0 \rho ^0$, $\pi ^+ \omega$, and $\pi ^0 \omega$
 (together with CP conjugate processes).
By restricting our analysis to these decays, we do not need to introduce
the additional parameters related to the singlet amplitudes.
Our omission of the strangeness-changing $B \to PV$ decays is deliberate, 
since such amplitudes seem to exhibit
 some anomalous behaviour already in the $B \to PP$ sector 
\cite{BF2000}.

The respective $B \to \pi \rho ,\pi \omega $ amplitudes are given by
(in sign convention used e.g. in ref.\cite{GR2004})
\begin{eqnarray}
\label{Amplpiplusrhominus}
A(B^0_d \to \pi ^+ \rho ^-)&=&-[\tilde{T_V}+ P_V]\\
\label{Amplpiminusrhoplus}
A(B^0_d \to \pi ^- \rho ^+)&=&-[\tilde{T_P}+ P_P]\\
\label{Amplpi0rhoplus}
A(B^+ \to \pi ^0 \rho ^+)&=&
-\frac{1}{\sqrt{2}}[\tilde{T_P}+\tilde{C}_V+P_P -P_V]\\
\label{Amplpiplusrho0}
A(B^+ \to \pi ^+ \rho ^0)&=&
-\frac{1}{\sqrt{2}}[\tilde{T_V}+\tilde{C}_P-P_P +P_V]\\
\label{AmplC1}
A(B^0_d \to \pi ^0 \rho ^0)&=&
-\frac{1}{2}[\tilde{C}_P+\tilde{C}_V-P_P -P_V]\\
\label{AmplC2}
A(B^+ \to \pi ^+ \omega)&=&
\frac{1}{\sqrt{2}}[\tilde{T}_V+\tilde{C}_P+P_P +P_V]\\
\label{AmplC3}
A(B^0_d \to \pi ^0 \omega)&=&
\frac{1}{2}[\tilde{C}_P-\tilde{C}_V+P_P +P_V]
\end{eqnarray}

where
\begin{equation}
\label{PPc}
P_{P(V)}\equiv P_{P(V),c}=A \lambda ^3 {\cal{P}}_{P(V),tc}\\
\end{equation}
and $\tilde{T}_{P(V)}$, $\tilde{C}_{P(V)}$ involve expressions similar to
Eqs (\ref{Btopipi2},\ref{Btopipi3}).

Following the arguments given in \cite{Lipkin} and used in previous discussions
\cite{RosnerLipkin}, we assume that $P_V=-P_P$, or, more precisely, that:
\begin{equation}
{\cal{P}}_{V,tc(u)}=
-{\cal{P}}_{P,tc(u)}.
\end{equation}

Then, 
defining $P_{P,u}$ and $P_{V,u}$ in analogy to Eq. (\ref{PPc})
\begin{equation}
P_{P(V),u}=A \lambda ^3 {\cal{P}}_{P(V),tu}
\end{equation}
and inserting ${\cal{P}}_{P,tc(u)}=\xi {\cal{P}}_{tc(u)}$ from Eq. (\ref{ksi}),
the amplitudes $\tilde{T}_{P(V)}$ and $\tilde{C}_{P(V)}$
may be written as
\begin{eqnarray}
\label{TV}
\tilde{T}_V= e^{i \gamma} (T_V-R_b P_{V,u})&=&
A \lambda ^3 R_b e^{i \gamma}({\cal{T}}_V+\xi {\cal{P}}_{tu}) \\
\label{TP}
\tilde{T}_P= e^{i \gamma} (T_P-R_b P_{P,u})&=&
A \lambda ^3 R_b e^{i \gamma}({\cal{T}}_P-\xi {\cal{P}}_{tu}) \\
\label{CV}
\tilde{C}_V= e^{i \gamma} (C_V+R_b P_{V,u})&=&
A \lambda ^3 R_b e^{i \gamma}({\cal{C}}_V-\xi {\cal{P}}_{tu})\\
\label{CP}
\tilde{C}_P= e^{i \gamma} (C_P+R_b P_{P,u})&=&
A \lambda ^3 R_b e^{i \gamma}({\cal{C}}_P+\xi {\cal{P}}_{tu})
\end{eqnarray}
with $P_{P,u}=\xi P_u=-P_{V,u}$.
In the above equations, 
the rightmost entries are given in the form completely analogous
to that used in ref. \cite{BurasFleischer}, with
$T_{V(P)}=A \lambda ^3 R_b {\cal{T}}_{V(P)}$ and
$C_{V(P)}=A \lambda ^3 R_b {\cal{C}}_{V(P)}$ 
involving the strong amplitudes ${\cal{T}}_{V(P)}$ and ${\cal{C}}_{V(P)}$
of colour-allowed  and
colour-suppressed  tree diagrams.

Since the amplitude $P_P=P_{P,c}=-P_V$ is known (Eq. (\ref{PPcvalue})),
from Eq. (\ref{Amplpiplusrhominus}) and the knowledge
of experimental asymmetries and branching ratios for the
 $B^0_d \to \pi ^+ \rho ^-$ decay one can determine the magnitude and
 (relative) phase of amplitude $\tilde{T}_V$. Then,
 using Eq. (\ref{TV}) with $P_{V,u}=-P_{P,u}$
  and the estimate $P_{P,u}=P_{P,c} P_u/P_c \approx
 -P_{P,c} xe^{i(\Delta-\theta)}/(R_b d) $  
  one can extract the tree amplitude $T_V$.
 A similar procedure applied to Eqs (\ref{Amplpiminusrhoplus},\ref{TP}) yields
 tree amplitude $T_P$.
 A subsequent use of Eqs (\ref{Amplpi0rhoplus},\ref{Amplpiplusrho0})
 should permit the determination of $C_V$ and $C_P$.
 The remaining three equations (\ref{AmplC1},\ref{AmplC2},\ref{AmplC3})
 provide additional constraints/check on the extracted values of
 colour-suppressed amplitudes.
 We now turn to the extraction of the relevant amplitudes from the data.
 
 \subsection{Extraction of tree amplitudes \mbox{\boldmath $T_V$} 
 and \mbox{\boldmath $T_P$}}
In analogy with Eq. (\ref{dtheta}), 
we first introduce the following parameters in the
$B\to PV $ sector:
\begin{eqnarray}
\label{TPfromTPtilde}
d_P e^{i\theta _P}&=&-e^{i \gamma} \frac{P_P}{\tilde{T}_P}
=-\frac{|P_P|}{|\tilde{T}_P|}e^{-i\delta _{\tilde{T}_P}}
=-\frac{P_{P,c}}{T_P-R_bP_{P,u}}\\
\label{TVfromTVtilde}
d_V e^{i\theta _V}&=&-e^{i \gamma} \frac{P_V}{\tilde{T}_V}
=+\frac{|P_P|}{|\tilde{T}_V|}e^{-i\delta _{\tilde{T}_V}}
=+\frac{P_{P,c}}{T_V+R_bP_{P,u}}.
\end{eqnarray}

For the CP-averaged branching ratio 
\begin{equation}
\label{BV}
\overline{{\cal{B}}}(B^0_d\to \pi ^+ \rho ^-)=
\frac{1}{2}[{\cal{B}}(B^0_d\to \pi ^+ 
\rho ^-)+{\cal{B}}(\bar{B}^0_d \to \pi ^- \rho ^+)]
\end{equation}
and the CP-asymmetry 
\begin{equation}
\label{AV}
A(B^0_d\to \pi ^+ \rho ^-)=
\frac{{\cal{B}}(\bar{B}^0_d \to \pi ^- 
\rho ^+)-{\cal{B}}(B^0_d\to \pi ^+ \rho ^-)}
{{\cal{B}}(B^0_d\to \pi ^+ 
\rho ^-)+{\cal{B}}(\bar{B}^0_d \to \pi ^- \rho ^+)}
\end{equation}
 one derives:
\begin{eqnarray}
\label{BCPV}
\overline{{\cal{B}}}(B^0_d\to \pi ^+ \rho ^-)&=&
\left[ 1+\frac{1}{d^2_V}-\frac{2}{d_V}\cos \theta_V \cos \gamma \right] 
|P_P|^2\\
\label{ACPV}
A(B^0_d\to \pi ^+ \rho ^-)&=&\frac{{\mbox{\large $\frac{4}{d_V}$}}\sin \theta_V \sin \gamma}
{1+{\mbox{\large $\frac{1}{d^2_V}$}}-
{\mbox{\large $\frac{2}{d_V}$}}\cos \theta_V \cos \gamma}
\end{eqnarray}
with (Eq. (\ref{PPcvalue}))
\begin{equation}
\label{PPvalue}
P_P=0.69 \pm 0.05.
\end{equation}

Solving Eqs (\ref{BCPV},\ref{ACPV}) one gets
\begin{eqnarray}
\label{tanThetaVkV}
\tan \theta _V &=& \frac{k_V}
{\sin \gamma (\cos \gamma \pm 
\sqrt{\cos ^2 \gamma +
l_V-1-k_V^2/\sin ^2 \gamma })}\\
\label{dVkV}
d_V & = & \frac{\sin \theta _V \sin \gamma}{k_V}
\end{eqnarray}
where we defined
\begin{eqnarray}
k_V &\equiv  &\frac{A(B^0_d\to \pi ^+ \rho ^-)
\overline{{\cal{B}}}(B^0_d\to \pi ^+ \rho ^-)}{4 |P_P|^2}\\
\label{lV}
l_V &\equiv & \overline{{\cal{B}}}(B^0_d\to \pi ^+ \rho ^-)/
|P_P|^2.
\end{eqnarray}
With the experimental values:
\begin{eqnarray}
\overline{{\cal{B}}}(B^0_d\to \pi ^+ \rho ^-)&=&10.1 ^{+2.1}_{-1.9}\\
A(B^0_d\to \pi ^+ \rho ^-)&=& -0.47 ^{+0.13}_{-0.14}
\end{eqnarray}
taken from \cite{HFAGsummer2004},  after neglecting the correlations
with $B^0_d\to \pi ^- \rho ^+$, one obtains the following two solutions:
\begin{eqnarray}
\label{V1sol}
d_V=d_{V,1}&\equiv & 0.206 ^{+0.025}_{-0.022}\\
\theta_V=\theta _{V,1} & \equiv & (-34 ^{+12}_{-15})^o
\end{eqnarray}
or
\begin{eqnarray}
\label{V2sol}
d_V=d_{V,2}&\equiv & 0.239 ^{+0.037}_{-0.032}\\
\theta_V=\theta _{V,2} & \equiv & (-139 ^{+17}_{-13})^o.
\end{eqnarray}

{{From the CP-averaged branching ratio 
$\overline{{\cal{B}}}(B^0_d\to \pi ^- \rho ^+)$ and asymmetry 
$A(B^0_d\to \pi ^- \rho ^+)$ defined in  analogy to Eqs (\ref{BV}, \ref{AV}),
using the experimental values
\begin{eqnarray}
\overline{{\cal{B}}}(B^0_d\to \pi ^- \rho ^+)&=&13.9 ^{+2.2}_{-2.1}\\
A(B^0_d\to \pi ^- \rho ^+)&=& -0.15 \pm 0.09
\end{eqnarray}
from \cite{HFAGsummer2004},
one similarly obtains:}}
\begin{eqnarray}
\label{P1sol}
d_P=d_{P,1}&\equiv & 0.174 ^{+0.019}_{-0.017}\\
\theta_P=\theta _{P,1} & \equiv & (-12 ^{+7}_{-8})^o
\end{eqnarray}
or
\begin{eqnarray}
\label{P2sol}
d_P=d_{P,2}&\equiv & 0.203 ^{+0.026}_{-0.023}\\
\theta_P=\theta _{P,2} & \equiv & (-166 ^{+9}_{-9})^o.
\end{eqnarray}

Further experimental constraints are given
by the parameters $S_{\rho \pi}$ and $\Delta S_{\rho \pi}$ 
extracted from the time-dependent studies of 
$(B^0_d,\bar{B}^0_d)\to\rho ^{\pm}\pi ^{\mp}$,
and providing information on the relative phases of effective tree
amplitudes
$\tilde{T}_V$ and $\tilde{T}_P$.
In terms of our amplitudes of 
Eqs (\ref{Amplpiplusrhominus},\ref{Amplpiminusrhoplus}) and their
CP-counterparts,
these parameters are expressed as follows:

\begin{eqnarray}
S_{\rho\pi}& =&(S_{+-}+S_{-+})/2\\
\Delta S_{\rho\pi}&=&(S_{+-}-S_{-+})/2
\end{eqnarray}
with
\begin{eqnarray}
S_{+- }&\equiv& \frac{2~{\rm Im}~\lambda^{+-}}{1+|\lambda^{+-}|^2}\\
S_{-+}&\equiv& \frac{2~{\rm Im}~\lambda^{-+}}{1+|\lambda^{-+}|^2}\\
\lambda ^{+-}&\equiv & e^{-2i\beta}\frac{A(\bar{B}^0_d\to\rho^+\pi^-)}
{A({B}^0_d\to\rho^+\pi^-)}=-e^{i(\theta_P-\theta_V-2\beta)}\frac{d_P}{d_V}
\frac{e^{-i\gamma}-d_Ve^{i\theta_V}}{e^{i\gamma}-d_Pe^{i\theta_P}}\\
\lambda ^{-+}&\equiv & e^{-2i\beta}\frac{A(\bar{B}^0_d\to\rho^-\pi^+)}
{A({B}^0_d\to\rho^-\pi^+)}=-e^{i(\theta_V-\theta_P-2\beta)}\frac{d_V}{d_P}
\frac{e^{-i\gamma}-d_Pe^{i\theta_P}}{e^{i\gamma}-d_Ve^{i\theta_V}}
\end{eqnarray}

{{The four pairs of solutions, i.e. $(P_1,V_1)$, $(P_1,V_2)$, $(P_2,V_1)$, 
and $ (P_2,V_2)$, give the four predictions for $S_{\rho\pi}$ and
$\Delta S_{\rho \pi}$ gathered in Table \ref{table0}.}}

\begin{table}[t]
\caption{ {Four predictions for $S_{\rho\pi}$ and
$\Delta S_{\rho \pi}$}}
\label{table0}
\begin{center}
\begin{tabular}{ccccc}
\hline
\rule[-6pt]{0pt}{20pt}  & case (a) & case (b)& case (c) & case (d) \\
&$(P_1,V_1)$ & $(P_1,V_2)$ & $(P_2,V_1)$ & $ (P_2,V_2)$\\
\hline
\rule[-6pt]{0pt}{24pt}$S_{\rho\pi}$  
& $-0.27^{+0.07}_{-0.04}$ & 
$-0.04^{+0.04}_{-0.04}$ & $0.00^{+0.04}_{-0.04}$ & $+0.32^{+0.04}_{-0.07}$ \\
\rule[-8pt]{0pt}{22pt} $\Delta S_{\rho \pi}$  
& $+0.38^{+0.25}_{-0.24}$ & 
$+0.73^{+0.16}_{-0.17}$ & $-0.73^{+0.19}_{-0.16}$ & $-0.38^{+0.22}_{-0.24}$\\
\hline
\end{tabular}
\end{center}
\end{table}





Since, according to the data \cite{HFAGsummer2004}:
\begin{eqnarray}
S_{\rho\pi}&=&-0.15\pm 0.13\\
\label{expDeltaSrhopi}
\Delta S_{\rho \pi}&=&+0.25 \pm0.13
\end{eqnarray}
cases (c) and (d) may be rejected.
Although case (a)  is clearly the best, case (b) cannot be ruled out:
the difference of the two determinations of $\Delta S_{\rho \pi}$ 
(case (b), Eq. (\ref{expDeltaSrhopi})) is consistent with 
zero at (slightly above) $2 \sigma$.

In both cases (a) and (b), we have
\begin{equation}
\label{TPtildetoPcvalue}
e^{-i\gamma }\tilde{T}_P/P_c=(3.68^{+0.46}_{-0.43})~e^{i (-168^{+8}_{-7})^o}.
\end{equation}

{{For $e^{-i\gamma }\tilde{T}_V/P_c$, the relative phase
$\delta \equiv {\rm Arg}(\tilde{T}_V/\tilde{T}_P)=\theta_P-\theta_V-180^o$, 
and the relative size $\left| {\tilde{T}_P}/{\tilde{T}_V}\right|={d_V}/{d_P}$ of tree amplitudes,
one finds the results gathered in Table \ref{table0a}.}}
\begin{table}[t]
\caption{{Relative sizes and phases of effective tree amplitudes}}
\label{table0a}
\begin{center}
\begin{tabular}{ccc}
\hline
\rule[-6pt]{0pt}{20pt}  & case (a) & case (b) \\
\hline
\rule[-6pt]{0pt}{24pt}$e^{-i\gamma }\tilde{T}_V/P_c $
& $(3.11^{+0.42}_{-0.39}) e^{i(+34^{+15}_{-12})^o}$ & 
$(2.68^{+0.45}_{-0.40}) e^{i(+139^{+13}_{-17})^o}$ \\
\rule[-8pt]{0pt}{22pt} $\delta  $
& $(-158 ^{+14}_{-17})^o$ & 
$(-53 ^{+19}_{-15})^o$ \\
\rule[-8pt]{0pt}{22pt}
$\left|
{\tilde{T}_P}/{\tilde{T}_V}\right|$&$1.18^{+0.20}_{-0.16}$& $1.37^{+0.26}_{-0.22}$\\
\hline
\end{tabular}
\end{center}
\end{table}
These solutions should be compared with $\delta \approx -22^o$ and
$|\tilde{T}_P/\tilde{T}_V| \approx 1.46 $ obtained 
in the favored fit of ref. \cite{GR2004}, and corresponding to our case (b).

One of the essential differences with ref. \cite{GR2004} is the fact that in our
approach the effective tree amplitudes $\tilde{T}_P$ and $\tilde{T}_V$ do not
correspond to the tree amplitudes $T_P$ and $T_V$ of the factorization 
picture. Instead, the effective amplitudes
$\tilde{T}_P$ and $\tilde{T}_V$
involve substantial corrections to the factorization terms,
due to the presence of the $P_{P,u}$ 
 ($P_{V,u}$ ) part of the penguin amplitude (Eqs (\ref{TV},\ref{TP})).
Using Eqs (\ref{TPfromTPtilde},\ref{TVfromTVtilde}) one can determine the
 tree amplitudes:
\begin{eqnarray}
\label{TPtoPcformula}
\frac{T_P}{P_c}&=&+\xi \left( R_b \frac{P_u}{P_c}-\frac{e^{-i\theta_P}}{d_P}
\right)\\
\label{TVtoPcformula}
\frac{T_V}{P_c}&=&-\xi \left( R_b \frac{P_u}{P_c}-\frac{e^{-i\theta_V}}{d_V}
\right)
\end{eqnarray}

From Eq.(\ref{TPtoPcformula}), using the estimate (see Eq.(\ref{PctoPuapprox}))
\begin{equation}
\label{estimRBPutoPc}
R_b\frac{P_u}{P_c}\approx - \frac{x}{d}e^{i(\Delta-\theta)}
\end{equation}
and Eq.(\ref{Deltamintheta2}), for both cases (a) and (b) one obtains:
\begin{equation}
\label{TPtoPcvalue}
\frac{T_P}{P_c}=(2.40^{+0.61}_{-0.61})e^{i(-157^{+15}_{-13})^o}.
\end{equation}
Similarly, from Eq.(\ref{TVtoPcformula}) one gets: 
\begin{eqnarray}
\label{aTVtoPc}
{\rm case~(a)}~~~~~~~~~~~~~~~~~~~\frac{T_V}{P_c}&=
&(2.25^{+0.60}_{-0.50})e^{i(+58^{+23}_{-20})^o~}~~~~~~~~~~~~~~~~~~~~~~~~~~\\
\label{bTVtoPc}
{\rm case~(b)}~~~~~~~~~~~~~~~~~~~\frac{T_V}{P_c}&=
&(3.91^{+0.71}_{-0.64})e^{i(+150^{+10}_{-12})^o}~~~~~~~~~~~~~~~~~~~~~~~~~~
\end{eqnarray}
For case (a) one finds $|T_V/T_P|=0.94^{+0.41}_{-0.29}$, while for case (b):
$|T_V/T_P|=1.63^{+0.63}_{-0.38}$. 
If the $B \to \rho$ and $B \to \pi$ formfactors
are similar, one expects (see \cite{GR2004}) 
that the ratio of $|T_V/T_P|$ should be approximately
equal to the
ratio of $f_{\pi}/f_{\rho}\approx 0.63 $, as $T_V$ ($T_P$) 
involves a weak current producing 
$\pi ^{\pm}$ ($\rho ^{\pm}$).
Thus, case (a) seems favoured again.

For both cases (a) and (b), however, one also estimates
from Eqs (\ref{TtoPcvalue},\ref{TPtoPcvalue}) 
that $|T_P/T| \approx 0.67 ^{+0.34}_{-0.25}$, which disagrees with the
simple expectation (c.f. \cite{GR2004}) of 
$|T_P/T| \approx f_{\rho}/f_{\pi}=1.59 $
(while $|\tilde{T}_P/\tilde{T}|=\xi d/d_P=1.91^{+0.84}_{-0.56}$).
Still, one has to keep in mind that the above estimates are based on
 Eq. (\ref{estimRBPutoPc}) which neglects terms
of order $C/T$.

\subsection{Extraction of colour-suppressed amplitudes 
\mbox{\boldmath $C_P$} and \mbox{\boldmath $C_V$}}

In analogy with Eqs (\ref{TPfromTPtilde},\ref{TVfromTVtilde}), 
we introduce the following parameters involving
colour-suppressed amplitudes $\tilde{C}_P$ and $\tilde{C}_V$:
\begin{eqnarray}
\label{yPdef}
y_P e^{i \Gamma _P}&=&-\frac{P_P}{\tilde{T}_V+\tilde{C}_P}e^{i\gamma}\\
\label{yVdef}
y_V e^{i \Gamma _V}&=&-\frac{P_V}{\tilde{T}_P+\tilde{C}_V}e^{i\gamma}.
\end{eqnarray}

The CP-averaged branching ratio for the $B^+ \to \pi ^+ \rho ^0$ decay and the
corresponding asymmetry are given by
\begin{eqnarray}
\label{eq119}
\bar{\cal{B}}(B^+\to \pi ^+ \rho ^0)&=& 
\left( 4+\frac{1}{y^2_P}+4 \cos \gamma \frac{\cos \Gamma _P}{y_P}\right) 
\frac{P^2_P}{2}\\
\label{eq120}
A(B^+ \to \pi ^+ \rho ^0)&=& - 4 \sin \gamma \frac{\sin \Gamma _P}{y_P}
\frac{1}{{4+{\mbox{\large $\frac{1}{y^2_P}$}}+4 
\cos \gamma {\mbox{\large $\frac{\cos \Gamma _P}{y_P}$}}}}.
\end{eqnarray}
Using the experimental numbers of
\begin{eqnarray}
\bar{\cal{B}}(B^+\to \pi ^+ \rho ^0)&=& 9.1 \pm 1.3\\
A(B^+ \to \pi ^+ \rho ^0)&=&-0.19 \pm 0.11
\end{eqnarray}
one finds two solutions:


sol.~(P1)
\begin{eqnarray}
y_P=y_{P,1}&\equiv &0.195 ^{+0.028}_{-0.024}\\
\Gamma = \Gamma _{P,1}&\equiv & (+23 \pm 14)^o
\end{eqnarray}
and

sol.~(P2)
\begin{eqnarray}
y_P=y_{P,2}&\equiv &0.149 ^{+0.015}_{-0.014}\\
\Gamma _P = \Gamma _{P,2}&\equiv &(+163^{+10}_{-11})^o.
\end{eqnarray}

From Eq.(\ref{yPdef}) one has:
\begin{equation}
\label{CtildePformula}
\frac{\tilde{C}_P}{P_c}=-\xi \left( \frac{1}{y_P}e^{-i\Gamma _P}
+\frac{1}{d_V}e^{-i \theta _V}
\right)e^{i \gamma}.
\end{equation}

Putting the estimates of $y_P$, $d_V$ etc. into Eq.(\ref{CtildePformula}), 
 one obtains
the values of $\frac{\tilde{C}_P}{P_c}e^{-i\gamma }$ given in Table
\ref{table1}.

\begin{table}[t]
\caption{Effective color-suppressed amplitudes $\tilde{C}_P$
from $B^+ \to \pi ^+ \rho^0$  decays}
\label{table1}
\begin{center}
\begin{tabular}{ccc}
\hline
\rule[-6pt]{0pt}{20pt} $e^{-i\gamma }\tilde{C}_P/P_c $ & case (a) & case (b) \\
\hline
\rule[-6pt]{0pt}{24pt}sol. (P1)  
& $(5.62^{+0.77}_{-0.84})~e^{i(- 175^{+11}_{-10})^o}$ & 
$(1.11^{+1.08}_{-0.69})~e^{i(-155^{+37}_{-57})^o}$ \\
\rule[-8pt]{0pt}{22pt} sol. (P2)  
& $(1.61^{+0.98}_{-0.65})~e^{i(- 17^{+39}_{-26})^o}$ & 
$(6.15^{+0.77}_{-0.87})~e^{i(-5^{+9}_{-9})^o}$ \\
\hline
\end{tabular}
\end{center}
\end{table}

For the $B^+ \to \pi^0 \rho ^+$ decays, one obtains formulas
completely analogous to (\ref{eq119}) and (\ref{eq120}), with
$y_P \to y_V$, and $\Gamma _P \to \Gamma _V$. Using the experimental branching
ratio and asymmetry:
\begin{eqnarray}
\bar{\cal{B}}(B^+\to \pi ^0 \rho ^+)&=& 12.0 \pm 2.0\\
A(B^+ \to \pi ^0 \rho ^+)&=&+0.16 \pm 0.13
\end{eqnarray}

one finds two solutions:
\begin{eqnarray}
{\rm sol.~ (V1)} ~~~~~~~~~~~~~~~~~~~~~~~~&&\nonumber\\
y_V=y_{V,1}&\equiv&0.165^{+0.025}_{-0.021}~~~~~~~~~~~~~~~~~~\\
\Gamma_V= \Gamma _{V,1}&\equiv&(-21^{+18}_{-19})^o
\end{eqnarray}
and
\begin{eqnarray}
{\rm sol.~ (V2)} ~~~~~~~~~~~~~~~~~~~~~~~~
&&\nonumber\\
y_V=y_{V,2}&\equiv &0.130^{+0.015}_{-0.013}~~~~~~~~~~~~~~~~~~\\
\Gamma _V =\Gamma _{V,2}&\equiv &(-163^{+15}_{-14})^o.
\end{eqnarray}

From Eq.(\ref{yVdef}) one has:
\begin{equation}
\label{tildeCVtoPc}
\frac{\tilde{C}_V}{P_c}=\xi \left( \frac{1}{y_V}e^{-i\Gamma _V}
+\frac{1}{d_P}e^{-i \theta _P}
\right)e^{i \gamma}.
\end{equation}

Putting 
the estimates of $y_V$, $d_P$ etc. into Eq. (\ref{tildeCVtoPc}), one obtains 
the values of $\frac{\tilde{C}_V}{P_c}e^{-i\gamma }$ 
given in Table \ref{table2}.
\begin{table}[t]
\caption{Effective color-suppressed amplitudes $\tilde{C}_V$
from $B^+ \to \pi ^0 \rho^+$  decays}
\label{table2}
\begin{center}
\begin{tabular}{cc}
\hline
\rule[-8pt]{0pt}{24pt}&$\frac{\tilde{C}_V}{P_c}e^{-i\gamma }$\\
\hline
\rule[-7pt]{0pt}{24pt}sol.~(V1)
&$7.53^{+0.84}_{-0.81}~e^{i(+ 17^{+11}_{-10})^o}$\\
\rule[-9pt]{0pt}{24pt}sol.~(V2)
&$2.47^{+1.17}_{-0.97}~e^{i(+ 117^{+28}_{-22})^o}$\\
\hline
\end{tabular}
\end{center}
\end{table}

By comparing the experimental branching ratio for $B^0_d \to
\pi ^0 \rho ^0$ and the bound on $B^0_d \to \pi^0 \omega$
 (from \cite{HFAGsummer2004}) with the predictions of all
combinations of entries in Tables \ref{table1} and
\ref{table2}, one finds that only cases (a,P2,V2) and (b,P1,V2) 
may be admitted. We shall refer to them as Solutions I and II respectively.
The corresponding predictions for the branching ratios
of $B^0_d \to \pi ^0 \rho ^0$ and $B^0_d \to \pi ^0 \omega$ 
are compared with the data in Table \ref{table3}.

{{Discrepancies with experiment observed in Table \ref{table3} suggest 
that the assumptions 
(in particular the SU(3) assumption of Section 3
and/or possibly Eq. (\ref{universalpenguins})), 
which lead to the value of $P_P$ given in Eq. (\ref{PPvalue}) 
thereby affecting the 
extracted size of colour-suppressed $B\to PV$ effective amplitudes, 
might not be wholy adequate. One needs here a way of 
estimating the size of $P_P$ in the strangeness-preserving sector,
which would be both sufficiently precise and
less assumption-dependent.}}

\begin{table}[t]
\caption{Branching ratios for $B^0_d \to \pi ^0 \rho^0$ and $B^0_d \to \pi ^0
\omega $ decays}
\label{table3}
\begin{center}
\begin{tabular}{cccc}
\hline
\rule[-8pt]{0pt}{24pt}&exp & Sol. I & Sol. II \\
\hline
\rule[-7pt]{0pt}{24pt}$B^0_d \to \pi ^0 \rho^0$
&$5.0 \pm 1.8$
&$1.68^{+0.83}_{-0.51}$&$0.86 ^{+0.90}_{-0.49}$\\
$B^0_d \to \pi ^0 \omega $
&$<1.2$
&$2.28 ^{+0.81}_{-0.65}$&$2.31 ^{+0.87}_{-0.80}$\\
\hline
\end{tabular}
\end{center}
\end{table}

Solution II, with $y_P \approx 0.195 $, 
is fully consistent with the information gained from the branching
ratio  ${\cal{B}}(B^+ \to
\pi ^+ \omega)=5.9 \pm 0.8$, which yields $y_P=0.20 \pm 0.02$. 
Solution I,
with $y_P \approx 0.149 $, agrees
with $y_P$ determined from $B^+ \to \pi ^+ \omega$ at $2 \sigma$. 
Thus, the  $B^+ \to \pi ^+ \omega$ branching ratio 
favors Solution II over Solution I.
We recall that it is just the opposite case with the values of
$S_{\rho \pi}$ and $\Delta S_{\rho \pi}$ which favor case (a)
(hence Solution I) over case (b) (Solution II) by $2 \sigma $.
Since  for the 
 $B^0_d \to \pi ^0 \rho^0$ branching ratio, 
 as Table \ref{table3} shows,
  the difference between experiment and theory 
  is $1.7 \sigma$ ($2.1 \sigma$) for Solution I (II),
one concludes 
that Solution I describes the data slightly better than
Solution II.

The colour-suppressed factorization amplitudes may be estimated from
\begin{eqnarray}
\label{CPtoPcexact}
\frac{C_P}{P_c}&=& -\xi \left[ 
\frac{1}{y_P}e^{-i \Gamma _P}+\frac{1}{d_V} e^{-i \theta _V}+R_b\frac{P_u}{P_c}
\right]\\
\label{CPapproxCtoT0}
&\approx&-\xi \left[ 
\frac{1}{y_P}e^{-i \Gamma _P}+\frac{1}{d_V} e^{-i \theta _V}
-\frac{x}{d}e^{i(\Delta-\theta)}
\right],
\end{eqnarray}

and
\begin{eqnarray}
\frac{C_V}{P_c}&=& \xi \left[ 
\frac{1}{y_V}e^{-i \Gamma _V}+\frac{1}{d_P} e^{-i \theta _P}+R_b\frac{P_u}{P_c}
\right]\\
\label{CVapproxCtoT0}
&\approx&\xi \left[ 
\frac{1}{y_V}e^{-i \Gamma _V}+\frac{1}{d_P} e^{-i \theta _P}
-\frac{x}{d}e^{i(\Delta-\theta)}
\right].
\end{eqnarray} 
The values of $C_P$ and $C_V$ obtained for both Solutions I and II assuming
$C/T=0$ (Eqs (\ref{CPapproxCtoT0},\ref{CVapproxCtoT0}))
are gathered in Table \ref{table4}.

\begin{table}[t]
\caption{Color-suppressed factorization amplitudes $C_P$ and $C_V$
obtained for $C/T=0$}
\label{table4}
\begin{center}
\begin{tabular}{ccc}
\hline
\rule[-6pt]{0pt}{22pt}  & Sol. I &  Sol. II \\
\hline
\rule[-6pt]{0pt}{22pt}$C_P$&${\rm Re}(C_P/P_c)=0.15^{+0.88}_{-0.86}$
&$C_P/P_c=2.40^{+1.09}_{-0.82} e^{i(-173^{+21}_{-24})^o}$\\
\rule[-8pt]{0pt}{20pt}&${\rm Im}(C_P/P_c)=-0.31^{+1.01}_{-1.01}$&\\
\hline
\rule[-8pt]{0pt}{24pt}$C_V$&\multicolumn{2}{c}{$C_V/P_c=2.05^{+1.39}_{-1.30} e^{i(+82^{+22}_{-24})^o}$}\\
\hline
\end{tabular}
\end{center}
\end{table}

Interestingly, with the central value of $|T_P/P_c|$ being $2.40$,
 Solution I is consistent with a small value of $C_P/T_P$,
 while for Solution II the $C_P/T_P$ ratio is of the order of 1.
On the other hand, given the central $T_V/P_c$ value of 2.25 (3.91) for
Solution I(II) respectively (Eqs (\ref{aTVtoPc},\ref{bTVtoPc})),
 it is Solution II for which $C_V/T_V$ seems to be smaller.
 One has to remember, however, 
 that in our calculations we used the values of $x$ and $d$
 determined from the averages of not fully consistent asymmetries in the $\pi
 \pi$ sector. 
Furthermore, our estimate of errors in the determination of 
 $P_c/P_u$ did not include the errors due to nonvanishing $C/T$. 
 As remarked earlier,
 for $|C/T| \approx 0.2 $ these corrections may increase the error
 of $P_c/P_u$ by 20\%, affecting the ensuing discussion correspondingly
 (see Section 5).
  
 \section{Extraction of $C/T$ and $P_c/P_u$}
 In the analysis performed so far, the ratio $C/T$ of the factorization
 amplitudes in $B\to \pi \pi$ decays has been assumed negligible.
 It turns out, however, that one can actually
 {\em determine} the value of $C/T$ directly from the data, provided
 one is willing to make an {{additional}} very plausible assumption.
 Namely, we observe that the amplitudes $C$, $T$ in $B \to \pi \pi$ decay
 and the amplitudes $C_V$, $T_V$ in $B \to \pi \rho, \pi \omega$ transitions 
 are due to the same process, namely 
 a decay of $b$ quark into a pion and a light quark.
 The  difference between the two colour-suppressed amplitudes $C$ and $C_V$
 (and between the two tree amplitudes $T$ and $T_V$) should be 
 due only to
 the fact that the amplitude for the recombination of the
 freshly produced light quark with the spectator quark 
 depends on whether the two recombine into a pseudoscalar (in $B \to \pi \pi$)
  or a vector meson (in $B \to \pi \rho, \pi \omega $). However,
 this dependence on the recombination
 amplitude should cancel in the ratios, i.e. in $C/T$ and $C_V/T_V$.
 Consequently, we may assume that
 \begin{equation}
 \label{CtoTequalCVtoTV}
 \frac{C}{T}=\frac{C_V}{T_V}.
 \end{equation}
We now recall Eq. (\ref{PctoPu}), which correlates the ratio $P_c/P_u$
with the size of $C/T$. We seek a similar connection for the
$B \to \pi \rho, \pi \omega$ sector. 
To this end, we observe that using the expressions (\ref{TP},\ref{CV}) 
for the effective
amplitudes in Eq. (\ref{yVdef}) we can write:
\begin{equation}
\label{TPtogetridof}
\frac{T_P}{P_c}+\frac{C_V}{P_c}=\xi \left(
2 R_b \frac{P_u}{P_c}+\frac{1}{y_V} e^{-i\Gamma_V}
\right).
\end{equation}
Now, Eqs (\ref{TPfromTPtilde},\ref{TVfromTVtilde}) may be rewritten as
\begin{eqnarray}
\label{TPanddP}
\frac{T_P}{P_c}&=&\xi \left( 
R_b \frac{P_u}{P_c}-\frac{1}{d_P}e^{-i\theta_P}
\right) \\
\label{TVanddV}
\frac{T_V}{P_c}&=&\xi \left( 
-R_b \frac{P_u}{P_c}+\frac{1}{d_V}e^{-i\theta_V}
\right).
\end{eqnarray}
From Eqs (\ref{TPtogetridof},\ref{TPanddP}) we determine
\begin{equation}
\label{CVandyVanddP}
\frac{C_V}{P_c}=\xi \left( 
R_b \frac{P_u}{P_c}+\frac{1}{y_V}e^{-i\Gamma_V}+\frac{1}{d_P}e^{-i\theta_P}
\right).
\end{equation} 
Dividing Eq. (\ref{CVandyVanddP}) by Eq. (\ref{TVanddV}) we obtain
\begin{equation}
\label{CVtoTVandPutoPc}
\frac{C_V}{T_V}=
\frac{R_b\mbox{\large $\frac{P_u}{P_c}$}
+\mbox{\large $\frac{1}{y_V}e^{-i\Gamma_V}$}+
\mbox{\large $\frac{1}{d_P}e^{-i\theta_P}$}}
{\mbox{\large $\frac{1}{d_V}e^{-i\theta_V}$}
-R_b\mbox{\large $\frac{P_u}{P_c}$}}.
\end{equation}
The above equation may be
rewritten in the form completely analogous to Eq. (\ref{PctoPu}), namely:
\begin{equation}
\label{PctoPuandCVtoTV}
\frac{P_c}{P_u}=-\left( 1+\mbox{\large $\frac{C_V}{T_V}$}\right)\frac{R_b }
{\mbox{\large $\frac{1}{d_P}e^{-i\theta_P}$}+\mbox{\large 
$\frac{1}{y_V}e^{-i\Gamma_V}$}-
\mbox{\large $\frac{C_V}{T_V}\frac{1}{d_V}e^{-i\theta_V}$}}.
\end{equation}
In the denominator above, the first two terms partially cancel.
By assuming that $C_V/T_V$ is so small that the third term may be neglected, we 
obtain a counterpart of the previous
 estimate of $P_c/P_u$ given in Eq. (\ref{PctoPuapprox}):
 \begin{equation}
 \label{PctoPufromBtoPV}
 \frac{P_c}{P_u}=-~\frac{R_b}{\mbox{\large $\frac{1}{d_P}e^{-i\theta_P}$}+\mbox{\large 
$\frac{1}{y_V}e^{-i\Gamma_V}$}} \approx (0.10 ^{+0.05}_{-0.04}) e^{i(63
^{+22}_{-28})^o},
 \end{equation}
which, despite the approximation involved, is consistent 
with Eq. (\ref{PctoPuapprox}), and thus with a large value of
$P_u$ as compared with $P_c$ (and a small value of $C/T$).

If $C/T$ is assumed equal to $C_V/T_V$,
Eqs (\ref{PctoPu},\ref{PctoPuandCVtoTV})
may be solved for $C/T$ with the final result:
\begin{equation}
\label{CtoTfinal}
\frac{C}{T}=\frac{\mbox{\large $\frac{1}{d_P}e^{-i\theta_P}$}+
\mbox{\large $\frac{1}{y_V}e^{-i\Gamma_V}$}
-\mbox{\large $\frac{x}{d}e^{i(\Delta-\theta)}$}}
{\mbox{\large $\frac{1}{d_V}e^{-i\theta_V}$}-\mbox{\large 
$\frac{1}{d}e^{-i\theta}$}}.
\end{equation}
Let us take the central values of the parameters 
and discuss the denominator first.
Solutions I and II differ in their values for the parameters of the pair 
($d_V$,$\theta_V$).
For Solution I (II), the first term in the denominator has 
an absolute value
of around $4.9$ ($4.2$). 
The second term
has an absolute value of around $1.9$. The sum of the terms in the denominator,
with phases taken into account, has an absolute value of $6.8$ ($4.3$)
 for Solution I (II) respectively.
As for the numerator, our previous 
considerations  uniquely determined the values of the three numerator terms.
In particular, the absolute value of the first term is equal to 
$1/d_P=1/d_{P,1}\approx 5.7$, that of 
the second term is $1/y_V=1/y_{V,2}\approx 7.7$, while for
 the third term it is equal to $x/d \approx 2.2$.
It is therefore non-trivial that with the central values of 
phases taken into account,
the sum of
these terms is not large and has the absolute value of around $3.2$, leading to
the central value of $|C/T|$ for Solution I being $0.47$.
The sum of the three terms in the numerator is most sensitive to the value 
of angle $\Gamma _V$. If $\Gamma _V$ is set at its $1 \sigma$ deviation value of
$-177^o$, the absolute value of the numerator becomes $1.3$ only.
For Solution I, the value of $|C/T|$ would then become equal to $0.2$.
When all of the errors are calculated, 
{{one obtains the values given in Table \ref{tableCT}.}}

\begin{table}[t]
\caption{{Extracted values of $C/T$}}
\label{tableCT}
\begin{center}
\begin{tabular}{ccc}
\hline
\rule[-6pt]{0pt}{20pt}  & Solution I & Solution II \\
\hline
\rule[-6pt]{0pt}{24pt}$C/T$ 
& $(0.47^{+0.33}_{-0.30})~e^{i(+47^{+24}_{-27})^o}$ & 
$(0.75^{+0.52}_{-0.49})~e^{i(-30 \pm 24)^o}$ \\
\hline
\end{tabular}
\end{center}
\end{table}



{{The determinations of $C_V/T_V=C/T$ given in Table \ref{tableCT}
 may be compared with
the central value of $C_V/T_V=0.91 e^{i 24^o}$$(0.52 e^{-i 68^o})$ 
for Solution I (II) obtained in
Section 4 for $C/T=0$.}}
Thus, the previously obtained central value of $C_V/T_V$ 
 gets significantly reduced (increased) 
 for Solution I (II). 
 Although in Solution II the central value of $|C/T|$ is now quite large
 it is also compatible with $|C/T|$
of order $0.25$. Better data are clearly
required. 

{{Other estimates of $C/T$ also lead to values of order 0.5. For example, in
ref. \cite{BurasFleischer} arguments in favor of $C/T = 0.5\times e^{i290^o}$
are given. Similarly, in their
 recent SU(3)-symmetric fit to all $B \to PP$ decays, 
 Chiang et al. \cite{Alietal}
 obtain the value  $|C/T| = 0.46^{+0.43}_{-0.30}$.
}}

With the central values of $|C/T|$ in Table \ref{tableCT}
significantly larger
than the expected value of around $0.25$,
the original estimate of  $P_c/P_u$, obtained in
Eq. (\ref{PctoPuapprox}) upon assuming $C/T=0$, could be
substantially affected.
Solving Eqs (\ref{PctoPu},\ref{PctoPuandCVtoTV}) for $P_c/P_u$, one obtains
\begin{equation}
\label{PctoPusamouzgod}
\frac{P_c}{P_u}=R_b~ d ~e^{i \theta}~\frac{\mbox{\large 
$1-\frac{\kappa}{d}~e^{-i \theta}$}}
{\mbox{\large $1-\frac{\kappa}{d_V}~e^{-i \theta _V}$ }},
\end{equation}
where
\begin{equation}
\kappa=\frac{1+x e^{i\Delta }}{\mbox{\large $\frac{1}{d_P}e^{-i \theta_P}
+\frac{1}{y_V}e^{-i\Gamma_V}+\frac{1}{d_V}e^{-i\theta_V}$}}.
\end{equation}
Numerically, 
for Solution I one finds:
\begin{equation}
\frac{P_c}{P_u}=(0.21^{+0.09}_{-0.06})~e^{i(44^{+19}_{-23})^o},
\end{equation}
which still bears resemblance to $(0.17^{+0.08}_{-0.05} ) 
~e^{i (16^{+21}_{-28})^o}$ of Eq. (\ref{PctoPuapprox}).

\noindent
 On the other hand, for Solution II one obtains:
\begin{equation}
\frac{P_c}{P_u}=(0.71^{+1.52}_{-0.66})~e^{i(+20^{+87}_{-55})^o},
\end{equation}
with error estimates so large that they admit small values for both
 $|P_c/P_u|$ and $|P_u/P_c|$. Again, there is here a strong dependence on
 $\Gamma_V$, with larger values of $|P_c/P_u|$ attained when $\Gamma_V$ is set
 at its $1\sigma$ deviation value of $-148^o$.

From Eqs 
(\ref{TPtoPcformula},\ref{CPtoPcexact},\ref{PctoPusamouzgod}), 
one can further determine the corresponding values of $C_P/T_P$. 
{{They are gathered in Table \ref{tableCPTP}.}}

\begin{table}[t]
\caption{{Extracted values of ${C}_P/T_{P}$}}
\label{tableCPTP}
\begin{center}
\begin{tabular}{ccc}
\hline
\rule[-6pt]{0pt}{20pt} & Solution I & Solution II \\
\hline
\rule[-6pt]{0pt}{24pt}${\rm Re} (C_P/T_P)$  
& $-0.24 ^{+0.23}_{-0.21}$ & 
$+0.40 ^{+0.38}_{-0.30}$ \\
\rule[-8pt]{0pt}{22pt}  ${\rm Im} (C_P/T_P)$
& $+0.03 ^{+0.36}_{-0.35}$ & 
$+0.00 ^{+0.28}_{-0.26}$ \\
${C_P}/{T_P}$ & $(0.25^{+0.31}_{-0.21}) ~e^{i(173^{+67}_{-66})^o}$ 
& $(0.40^{+0.40}_{-0.26}) ~e^{i(0^{+36}_{-46})^o}$ \\
\hline
\end{tabular}
\end{center}
\end{table}

{{ From Tables \ref{tableCT} and \ref{tableCPTP} we see that}} 
it is Solution I which prefers 
smaller central values of both $C_V/T_V=C/T$ and $C_P/T_P$.
With present errors, however, both Solutions I and II are still compatible with 
$|C/T|$ and $|C_P/T_P|$ of around 0.25.

For completeness, we have also calculated the ratio $|T_V/T_P|$ obtaining
$0.96^{+0.19}_{-0.18}$ ($0.88^{+0.37}_{-0.26}$) for Solution I (II) 
respectively
(to be compared with the value of $f_{\pi}/f_{\rho}\approx 0.63$
expected in \cite{GR2004}).

\section{Summary}
In this paper we performed a joint analysis of the $B \to \pi \pi$
and $B \to \pi \rho, \pi \omega$ decays with the aim of studying the
effects of the presence of two independent superpositions of penguin
amplitudes on the possible values of colour-suppressed and tree
factorization amplitudes. 
{{Our analysis assumes that the formation of the final PP or PV pair is
independent of the penguin transition occurring before that formation takes
place. This constitutes a crucial assumption of our approach.}}
The analysis yields two sets of solutions for the effective colour-suppressed
($\tilde{C}_V,\tilde{C}_P$) and tree ($\tilde{T}_P,\tilde{T}_V$)
amplitudes in the $B \to \pi \rho, \pi \omega$ transitions,
 with one solution weakly
favoured over the other one.

Assuming the $C/T$ ratio in $B \to \pi \pi$ to be negligible,
we estimated the ratio of the two superpositions of penguin amplitudes, 
using it subsequently to
determine the values of $C_V,T_V,C_P,T_P$ from the data. 
This procedure yielded two sets of numerical estimates for 
$C_P/T_P$ and $C_V/T_V$.

By imposing the condition of equality for the ratios of $C/T$ and $C_V/T_V$
we determined the value of $C/T$ directly from the data.
The two solutions obtained are compatible both with a 
value of $|C/T|$ of around 0.25 {{and with the estimates from literature 
yielding $|C/T| \approx 0.5$}}, 
with errors still of the order of $0.3 - 0.4$.
The corresponding solutions for $P_c/P_u$ 
and $C_P/T_P$ have been given as well.
One of the solutions is preferred as it yields smaller central values of
both $C/T=C_V/T_V$ and $C_P/T_P$.
Discrimination between the solutions, and a
more precise determination of $C/T$, require better data. 
{{When such data become available, a
well-defined
value may be extracted for $C/T$ along the lines similar to those
presented here
and compared with expectations and other estimates, providing us with
more information on the $C/T$ ratio and the expected connection between
 penguin amplitudes in $B \to \pi \pi$ and $B \to \pi \rho, \pi \omega$ decays.}}

This work was supported in part 
by the Polish State Committee for Scientific
Research (KBN) as a research project over the period of 
2003-2006 (grant 2 P03B 046 25).

\vfill

\vfill


\begin{thebibliography}{99}
\bibitem{BurasFleischer} A. J. Buras, R. Fleischer, S. Recksiegel, and F.
Schwab, Phys. Rev. Lett. 92, 101804 (2004); A. J. Buras, R. Fleischer, S. Recksiegel, and F.
Schwab, Nucl.Phys.B697, 133 (2004).
\bibitem{bb} B. Aubert {\em et al.} [BaBar], Phys. Rev. Lett. 91, 241801 (2003);
 K. Abe {\em et al.},
[Belle], Phys. Rev. Lett. 91, 261801 (2003).
\bibitem{Alietal}A. Ali, E. Lunghi, and A. Ya. Parkhomenko, Eur. Phys. J.
C36,183 (2004); C.-W.Chiang, M. Gronau, J. L. Rosner, and D. A. Suprun, 
Phys. Rev. D70, 034020 (2004).
\bibitem{BURAS2004end} A. J. Buras, R. Fleischer, S. Recksiegel, and F.
Schwab, hep-ph/0410407; hep-ph/0411373.
\bibitem{LZ20022003} P. \L{}ach, P. \.Zenczykowski, Phys. Rev. D66,
054011 (2002); P. \.Zenczykowski, P. \L{}ach, Phys. Rev.
D69, 094021 (2004) .
\bibitem{Z2004} P. \.Zenczykowski, Phys.
Lett. B590, 63 (2004).
\bibitem{Cot} N. de Groot, W. N. Cottingham, I. B. Whittingham, Phys. Rev. D68,
 113005 (2003); W. N. Cottingham, I. B. Whittingham, F. F. Wilson,
hep-ph/0501040.
\bibitem{GR2004} C.-W. Chiang, M. Gronau, Z. Luo, J. L. Rosner, and D. A.
Suprun, Phys. Rev. D69, 034001 (2004).
\bibitem{HFAGsummer2004}HFAG \verb+
www.slac.stanford.edu/xorg/hfag/triangle/ichep2004/index.shtml+
\bibitem{Lipkin} H. J. Lipkin, Phys. Rev. Lett. 46, 1307 (1981); Phys. Lett.
B254, 247 (1991); 415, 186 (1997); 433, 117(1998).
\bibitem{RosnerLipkin} M. Gronau and J. L. Rosner, Phys. Rev. D61, 073008
(2000); C. W. Chiang and J. L. Rosner, Phys. Rev. D65, 074035 (2002).
\bibitem{BF2000} A. J. Buras, R. Fleischer, Eur. Phys. J. C16, 97 (2000).
\end{thebibliography}
\end{document}